\documentstyle[11pt]{article}

\begin{document}
\pagestyle{empty}
\begin{flushright}
{BROWN-HET-942} \\
{April 1994}
\end{flushright}
\vspace*{5mm}
\begin{center}
{\bf Singularity-Free Two Dimensional Cosmologies\footnote{Work supported in
part by the US Department of Energy under Grant DE-FG0291ER40688.}} \\
[10mm]
R. Moessner\footnote{e-mail address moessner@het.brown.edu} and M.
Trodden\footnote{e-mail address mtrodden@het.brown.edu}\\
[10mm]
Department of Physics \\
Brown University \\
Providence RI. 02912. \\
USA. \\[2cm]

{\bf Abstract}
\end{center}
\vspace*{3mm}
We present a class of theories of two dimensional gravity which admits
homogeneous and isotropic solutions that are nonsingular and asymptotically
approach a FRW matter dominated universe at late times. These models are
generalizations of two dimensional dilaton gravity and both vacuum solutions
and those including conformally coupled matter are investigated. In each case
our construction leads to an inflationary stage driven by the gravitational
sector. Our work comprises a simple example of the `Nonsingular Universe'
constructions of ref. [1]. \\

\newpage\setcounter{page}{1}\pagestyle{plain}

\section{Introduction}
\smallskip
\par
The singularity theorems of Penrose and Hawking\cite{P&H} pose serious physical
and philosophical problems concerning the behaviour of physics in certain
regions of spacetime. Taken at face value the theorems tell us that the
manifolds described by General Relativity (GR) are generically geodesically
incomplete - there are always geodesics that cannot be continued for arbitrary
values of their affine parameter. This conclusion is highly unsatisfactory for
a number of reasons.

Firstly, the existence of singularities limits the predictive power of any
physical theory for which the manifold is a background. Particles reaching the
singularity cease to exist, carrying with them vital information about the
particle's history.

A second problem involves the fact that, as experience with `physical'
solutions to GR shows, the existence of a singularity is often accompanied by
the unbounded increase of one of the curvature invariants of the theory.
Examples of this include the infinite Weyl tensor at the centre of a
Schwarzschild black hole and the `big bang' singularity encountered in
cosmological models. In fact it can be shown that if the singularity is reached
on a timelike curve in a globally hyperbolic spacetime then the Riemann tensor
$R_{\mu\nu\rho\sigma}$ is unbounded\cite{CC 75}.

Finally, in connection with both of the above there is the problem of the
quantum information loss paradox. This is associated with the evolution of pure
quantum states into mixed states during the evaporation of black holes.

It is a common belief that when we have a complete quantum theory of gravity
the singularities predicted for GR will not occur. The appearance of
singularities should be seen as a signal of the breakdown of GR and an
indication of the need for a new theory.

Since there is at present no theory of quantum gravity we shall attempt a more
modest goal. We shall implement the `Limiting Curvature Construction'
(LCH)\cite{GMF 88} in a class of two dimensional theories of gravity. By doing
so we shall obtain theories to which all solutions are singularity free and the
only curvature invariant, the Ricci scalar $R$, is bounded. Thus, our key
assumption is that those processes that may be possible for curing the
`sickness' of GR are representable at the classical level in a modified theory
of gravity.

The models we will consider are generalizations of two-dimensional {\it Dilaton
Gravity} which is obtained from the low energy limit of string theory\cite{CGHS
92} and is a theory of gravity coupled to a scalar field, the {\it dilaton}. We
generalize these models by allowing an arbitrary coupling between the fields
and a general potential for the scalar (which we still refer to as the
dilaton). In a theory derived from string theory such a scalar field would be
dynamical and the analysis of the model would be quite complicated. Here,
however,  we treat the scalar as a Lagrange multiplier.

Nonsingular black hole solutions to this theory were studied in\cite{TMB 93}
and so, in this paper, we shall concentrate on cosmological solutions to the
theory. Cosmological solutions to 1+1 dimensional dilaton gravity have been
studied by other authors\cite{MNY 93} but our goals differ from theirs in that
we wish to study generalizations of that theory with emphasis on the possible
removal of spacetime singularities.

The outline of this paper is as follows. In section 2 we review the model of
generalized dilaton gravity and obtain the field equations. In section 3 we
explicitly search for homogeneous cosmological solutions and show that our
construction yields theories with a `natural' initial inflationary stage
evolving into a FRW matter dominated universe. All manifolds described by this
theory are singularity-free. In section 4 we investigate the effect of
including a conformally coupled scalar field into the theory and show that this
does not affect our ability to construct nonsingular cosmological theories.
Finally in section 5 we conclude and speculate on the implications of our
results.

\section{Two Dimensional Dilaton Gravity}

Here we shall describe the two dimensional models that we have investigated.
Given that the model of dilaton gravity discussed in the previous section
appears to be important for physics we shall study general two dimensional
models describing a massless scalar field (which we shall still refer to as the
dilaton) coupled to gravity. Our aim is to identify classes of theories to
which all solutions are nonsingular. In this way we hope to pinpoint those
features of the models which may be important for gravitational physics if
nonsingular behaviour is manifested at the classical level. We shall present
the model and derive the general field equations in preparation for the
implementation of our construction in the next section.

Our starting point is the most general renormalizable lagrangian for gravity
coupled to a scalar field in 1+1 spacetime dimensions.

\begin{equation}
{\cal L} = \sqrt{-g} [D(\phi)R + G(\phi)(\nabla\phi)^2 + H(\phi)]
\end{equation}
with action
\begin {equation}
S[{\bf g},\phi] = \int {\cal L} \, d^2x
\end{equation}
where $D(\phi)$ is the coupling of the dilaton to gravity, $G(\phi)$ is just a
conformal prefactor for the dynamical terms for the dilaton and $H(\phi)$ is a
potential for the dilaton.

We are free to perform a conformal or Brans-Dicke transformation on this
Lagrangian\cite{W&B 81}\cite{B&O 91} via

\begin{equation}
g_{\mu\nu}^{ old} = e^{2\tau(\phi)} g_{\mu\nu}^{ new}
\end{equation}
If we now require that

\begin{equation}
\frac{d\tau}{d\phi}\frac{dD}{d\phi} = -G(\phi)
\end{equation}
then we may express the Lagrangian in the form

\begin{equation}
{\cal L} = \sqrt{-g}\, (V(\phi) + D(\phi)R)
\end{equation}
where the new potential is given by $V(\phi) = H(\phi)e^{2\tau(\phi)}$. This is
the fundamental Lagrangian which we shall consider. In particular note that it
is the metric appearing in this expression that governs the motion of test
particles. Note that in principle we may eliminate $\phi$ from this Lagrangian
and thus obtain a higher derivative theory describing the gravitational field

\begin{equation}
{\cal L}^{ new} = {\cal F}(R)
\end{equation}

Now consider the equations of motion arising from the variation of the action
with Lagrangian as in equation (5). Variations with respect to the scalar field
yield

\begin{equation}
\frac{\partial V}{\partial \phi}(\phi) =
 - \frac{\partial D}{\partial \phi}(\phi) R
\end{equation}
Now consider variations with respect to the metric
\begin{eqnarray}
\delta S & = & \int d^{2}x \ [(V(\phi) + D(\phi)R)\, \frac{1}{2}\,
               g_{\alpha\beta} \sqrt{-g}\, \delta g^{\alpha\beta}  +
               \nonumber \\
         &   &  \sqrt{-g}\, (-D(\phi)R_{\alpha\beta}\delta g^{\alpha\beta}-
               D(\phi)\nabla^{2} g_{\alpha\beta} \delta g^{\alpha\beta} +
               D(\phi) \nabla_{\alpha}\nabla_{\beta} \delta
               g^{\alpha\beta}) ]
\end{eqnarray}
Integrating by parts twice, ignoring surface terms and requiring $\delta S = 0$
gives

\begin{equation}
-D(\phi)R_{\alpha\beta} + (\nabla_{\alpha}\nabla_{\beta} - g_{\alpha\beta}
\nabla^{2})D(\phi) + \frac{1}{2} V(\phi)g_{\alpha\beta} +
\frac{1}{2} D(\phi)Rg_{\alpha\beta} = 0
\end{equation}
Finally we note that in two dimensions $R_{\alpha\beta}=g_{\alpha\beta}R/2$
which puts our field equations in the form

\begin{equation}
V(\phi)g_{\alpha\beta} = 2(\nabla^{2}g_{\alpha\beta} -
\nabla_{\alpha}\nabla_{\beta})D(\phi)
\end{equation}
Equations (7) and (10) are our basic equations. In particular, (7) is the
constraint equation due to the dilaton and it is through this that we shall
force our theories to have limiting curvature and to be singularity free. In
ref.~6 nonsingular static black hole solutions to this model were constructed.
Here we seek to address the case when the spacetime is homogeneous and
isotropic. To be exact we shall search for singularity free cosmological
models.

\section{Cosmologies in Two Dimensions}
In this section we shall construct a vacuum model for which all homogeneous and
isotropic solutions are nonsingular and approach a matter dominated
compactified FRW spacetime at late times. Note that in particular these
spacetimes exhibit no big-bang singularity.

We begin by making a redefinition of $\phi$ so that $D(\phi)=\phi^{-1}$. We are
perfectly free to do this and the choice is made for calculational simplicity.
Since we are looking for spatially homogeneous solutions we shall use the
ansatz
\begin{equation}
 g_{\mu \nu}=diag(-1,a^{2}(t))
\end{equation}
for the metric. With these choices the field equations become
\begin{equation}
 V(\phi)-2\frac{\dot{a}}{a}\frac{\dot{\phi}}{\phi^{2}} = 0
\end{equation}
and
\begin{equation}
 V(\phi)-2\frac{\ddot{\phi}}{\phi^{2}}+4\frac{{\dot{\phi}}^{2}}{\phi^{3}}=0
\end{equation}
where dot denotes differentiation with respect to time. Also our constraint
equation (7) becomes
\begin{equation}
 \phi^{2}\frac{\partial V}{\partial \phi}(\phi)=R
\end{equation}
We may combine (12) and (13) and, for $d\phi^{-1}/dt \neq 0$, integrate to
obtain the scale factor $a(t)$
\begin{equation}
a(t)=-\gamma\frac{\dot{\phi}}{\phi^{2}}
\end{equation}
where $\gamma > 0$ is an arbitrary integration constant. Substituting into (12)
we have
\begin{equation}
 V(\phi)=-\gamma^{-1}2{\dot a}
\end{equation}
Now, our construction consists of the following steps. Firstly we shall require
that the scale factor approach that of a matter dominated FRW spacetime at late
times. We shall then solve for the asymptotic behaviour of the potential in
this region. We then demand that the curvature be bounded at early times. This
yields a particular form for the scale factor which allows us also to solve for
the asymptotic form of the potential in the early universe. Finally we shall
choose a well-behaved interpolating potential and use this to solve for the
general scale factor for our model.

At late times, $t \rightarrow \infty$, we shall impose $a(t)=\nu t^{2/3}$,
where $\nu > 0$ is a constant. Integration of (15) yields
\begin{equation}
 \phi(t)=\frac{5}{3}\frac{\gamma}{\nu} t^{-5/3}
\end{equation}
up to an integration constant. From (16) the late time form of our potential is
given by
\begin{equation}
 V(\phi)=-\frac{4}{3}\frac{\nu}{\gamma}t^{-1/3}=-\alpha \phi^{1/5}
\end{equation}
where $\alpha$ is a constant constructed from the other constants we have
introduced.

Now, at early times, $ t \rightarrow - \infty$, we require the curvature, R, to
be bounded, $ R=2\beta^2$ say. By (14) this implies
\begin{equation}
 V(\phi)=\frac{-2\beta^2}{\phi}
\end{equation}
where $\beta >0$ is a constant. Then (13) becomes a harmonic oscillator
equation for $\phi^{-1}$ giving
\begin{equation}
 \phi(t)=\frac{1}{a_{1} e^{\beta t}+a_{2} e^{-\beta t}}
\end{equation}
with $a_{1}$ and $a_{2}$ arbitrary constants. Further, (15) can be rewritten as
\begin{equation}
a(t)=\gamma \frac{d}{dt}(\phi^{-1})
\end{equation}
So for various choices of $a_{1}$ and $a_{2}$ we can obtain an expanding de
Sitter solution with Hubble constant $H=2\beta^2$, ($a_2=0$), a contracting de
Sitter solution ($a_1=0$) or a de Sitter bounce ($a_1=-a_2$).

Note that we have obtained an initial inflationary phase by bounding R, the
only curvature invariant in 2 dimensions, and thus our solutions are
nonsingular (both in the sense of avoiding infinite curvature and having a
geodesically complete manifold). It is an attractive feature of this and other
higher derivative theories of gravity that inflation\cite{AG 81}, which
provides a solution to  the homogeneity, horizon and flatness problems in
cosmology, can be the result of the action of gravity alone and needs no matter
fields to drive it.

Now to find a model close to a compactified version of our real Universe, we
shall search for a solution which has a de Sitter expansion at early times and
evolves into a `matter-dominated' FRW universe at late times. To this end we
choose
\begin{displaymath}
  V(\phi) \rightarrow \frac{-2\beta^2}{\phi}\;\;\;\;{\rm as} \;\;\;\;\;\phi
\rightarrow \infty
\end{displaymath}
\[   V(\phi) \rightarrow -\alpha \phi^{1/5}  \;\;\;\;{\rm as} \;\;\;\;\;\phi
\rightarrow 0 \]
A possible choice for the interpolating potential is
\begin{equation}
 V(\phi)=-\frac{\alpha \phi^{1/5}}{1+\alpha (2\beta^2)^{-1} \phi^{6/5}}
\end{equation}
Explicit solution for the scale factor is not possible but it is only
neccessary to show that our metric is well behaved as it interpolates between
its asymptotic limits.
{}From (13) we have (in terms of $D \equiv \phi^{-1}$)
\begin{equation}
 2 \ddot{D} = \frac{\alpha D}{D^{6/5}+ b}
\end{equation}
where $b=\alpha/2\beta^2$. This may be integrated to give
\begin{equation}
 \dot{D}^{2}= \alpha f(D)
\end{equation}
where
\begin{eqnarray}
 f(x) & = & \frac{5}{4}x^{4/5}
-\frac{5}{12}b^{2/3}\ln(x^{4/5}-b^{1/3}x^{2/5}+b^{2/3}) \nonumber \\
     &   & - \;\; \frac{5}{2 \sqrt{3}}b^{2/3}
\arctan(\frac{2x^{2/5}-b^{1/3}}{\sqrt{3}b^{1/3}}) \nonumber \\
     &   & + \;\; \frac{5}{6} b^{2/3}\ln(x^{2/5}+b^{1/3})
-\frac{5}{4\sqrt{3}}b^{2/3}\arctan(\sqrt{3})
\end{eqnarray}
The scale factor, $a(t)$, may be expressed implicitly using (15) as
\begin{equation}
a(D)=\gamma \dot{D} = \gamma \sqrt{\alpha f(D)}
\end{equation}
where we have chosen the positive square root for consistency. In Figure~$1$ we
plot $a(D)$ for $\alpha=2\beta^2=\gamma= 1$. The analysis is not qualitatively
sensitive to the value of these positive constants.

It can be seen that a(D) is a real, positive, monotonically increasing function
of D for $ D>0$. Further, when the universe emerges from an expanding de Sitter
phase,
$\frac{\partial}{\partial{D}}\dot{D}=\ddot{D}/\dot{D}>0$ for all $D>0$ and $D$
is positive for all time because $ D \rightarrow 0^{+}$ as $ t \rightarrow
-\infty$ and $ \dot{D}(D)>0$ for all D and so we can use the fact that $
\dot{D}>0 \;\;\forall \;t$ to assert that the scale factor is also a
monotonically increasing function of time. Our construction is therefore
complete.

In order to check the above results we also studied numerically the case of a
contracting universe to ensure that the model made a smooth transition between
the two asymptotic regions. Beginning with final conditions in the $t
\rightarrow \infty$, matter dominated region we evolved the equations backwards
to the initial de Sitter solution at $t \rightarrow -\infty$. Our numerical
results confirmed the above analysis.

Note that, although $\phi \rightarrow \infty$ as $t \rightarrow -\infty$ this
is in fact a coordinate singularity. The coordinates $t,x$ which we have chosen
only cover half of de-Sitter space. If we consider de-Sitter space as a
hyperboloid embedded in a flat three-dimensional space then we may define new
coordinates which describe the whole of the space\cite{H&E}. In these
coordinates it is clear that the point $t=-\infty$ is a coordinate singularity
and that the space is geodesically complete since geodesics may be smoothly
continued through the point $t=-\infty$..

To summarize, we have shown that the class of theories with potentials
satisfying the asymptotic conditions (18), (19) have all homogeneous isotropic
solutions which are nonsingular. Further, in bounding the curvature at early
times our solutions have a `natural' inflationary stage driven by the
gravitational sector. Figure~2 shows a log-log plot of the scale factor
demonstrating the interpolation between the behaviours of the asymptotic
regions.

We shall now turn our attention to modifications to the vacuum theory.

\section{Inclusion of Conformal Matter}
Now that we have found a class of theories for which all vacuum cosmological
solutions are singularity free we wish to extend our analysis to consider the
effect of mattter on our construction.

In these two dimensional models the requirement of quantum consistency of the
theory is equivalent to the conformal invariance of the Lagrangian. We plan to
include matter using a scalar field and since the dynamical terms imply that a
scalar field must transform trivially under a conformal transformation, we may
not include a simple, polynomial potential in the Lagrangian. Thus, by far the
most simple choice is to conformally couple a scalar field into the theory
using just the dynamical terms. We shall represent this matter by $\sigma$ and
shall demonstrate that the presence of this field does not prevent us from
constructing singularity free theories. We shall again restrict ourselves to
search for theories with homogeneous, isotropic cosmological solutions.

Our action is now

\begin {equation}
S[{\bf g},\phi;\sigma] = \int  \ ({\cal L}_g + {\cal L}_m)\ d^2x
\end{equation}
where ${\cal L}_g$ represents the gravitational sector of the theory given by
the previous Lagrangian (26) and ${\cal L}_m$ represents the Lagrangian for the
matter described by the conformally coupled scalar field $\sigma$.

\begin{equation}
{\cal L}_m = \lambda^2(\partial_\mu \sigma \partial^\mu \sigma) \sqrt{-g}
\end{equation}

The equations of motion derived from this action are the constraint equation
(14), the equation of motion for $\sigma$

\begin{equation}
\nabla^2 \sigma = 0
\end{equation}
and the modified field equations for the metric which, with the ansatz (11) and
$D(\phi)=\phi^{-1}$, take the form
\begin{equation}
\frac{{\ddot \phi}}{\phi^2} - 2\frac{{\dot \phi}^2}{\phi^3} -
\frac{1}{2}V(\phi) -\frac{\lambda^2}{2}{\dot \sigma}^2 = 0
\end{equation}
and
\begin{equation}
\frac{{\dot \phi}}{\phi^2} \frac{{\dot a}}{a} - \frac{1}{2}V(\phi) +
\frac{\lambda^2}{2}{\dot \sigma}^2= 0
\end{equation}
Since we seek solutions that depend only on time, we may solve (29) exactly to
give

\begin{equation}
{\dot \sigma} = \frac{\alpha}{a}
\end{equation}
Substituting this into our field equations and combining we obtain

\begin{equation}
\left(\frac{{\dot \phi}}{\phi^2}\right)^. - \frac{{\dot
\phi}}{\phi^2}\frac{{\dot a}}{a} - \frac{\lambda^2 \alpha^2}{a^2} = 0
\end{equation}
and

\begin{equation}
\left(\frac{{\dot \phi}}{\phi^2}\right)^. + \frac{{\dot
\phi}}{\phi^2}\frac{{\dot a}}{a} - V(\phi) = 0
\end{equation}
Now, as before, let us search for solutions that asymptotically approach a
matter-dominated FRW universe with scale factor given by $a(t)=\nu t^{2/3}$. In
this region it is straightforward to solve for the evolution of the dilaton and
we obtain

\begin{equation}
\phi(t) = \left[\frac{3\lambda^2 \alpha^2}{2\nu ^2} t^{2/3} - \frac{3}{5}\delta
t^{5/3}\right]^{-1}
\end{equation}
where $\delta$ is a constant of integration. This expression is valid for late
times, $t\rightarrow \infty$.

For $\delta\neq 0$ the second term in eq.(35) dominates. Eq.(34) can then be
solved for the potential at late times as
\begin{equation}
V(\phi)\rightarrow \;-\mu\phi^{1/5} \;\;\;\rm{as} \;\;\;\;t\;\rightarrow\infty
\;\;\;\;\;(\phi\rightarrow 0)
\end{equation}
where $\mu= (3\delta/5)^{1/5}(4\delta/3)$ .

At early times, ${\rm t} \rightarrow 0$, we shall impose the condition that the
curvature remain bounded, $R=2 \beta^{2}$ , where $2 \beta^2$ is a constant
representing the limiting curvature. Equation (14) then implies
\begin{equation}
 V(\phi)=\frac{-2\beta^2}{\phi}  \;\;\;\;\; {\rm as} \;\;\;\; t \rightarrow 0
\end{equation}
For the purposes of this paper we shall concentrate on the solution where the
universe emerges from a de Sitter bounce at $t=0$ since this case turns out to
be the easiest to treat analytically. Thus we write the scale factor at early
times as
\begin{equation}
a(t) =  \cosh(\beta t)
\end{equation}
where we have chosen $a(0)=1$ without loss of generality.
We can combine (33) and (34) into the following two equations, written in
terms of $D = \phi^{-1}$
\begin{equation}
2 \ddot D + V(D) +\frac{\kappa^{2}}{a^{2}} = 0
\end{equation}
\begin{equation}
2 \dot D  \frac{\dot a}{a} + V(D) - \frac{\kappa^{2}}{a^{2}} = 0
\end{equation}
where $\kappa^{2}=\lambda^{2} \alpha^{2}$. Equation (39) with $a(t)$ from (38),
linearized about $t=0$ , admits the solution
\begin{equation}
D(t) = \frac{\kappa^{2}}{2\beta^{2}} -2c \cosh{(\beta t)}
\end{equation}
where $c$ is an integration constant. Provided that
\begin{equation}
c=\frac{\kappa^{2}}{2\beta^{2}}
\end{equation}
(40) is also satisfied by this solution to linear order in $t$.
For $c \ll 1$, $D(t)$ tends to the small constant $-c$ as $t \rightarrow 0$ ,
so that the limit $t \rightarrow 0$ corresponds to $|\phi| \gg 1$.
To connect the two asymptotic regions of the de Sitter bounce and matter
domination, we can then find a form of the interpolating potential similar to
the one  used in the vacuum model
\begin{equation}
V(\phi)=-\frac{\mu\phi^{1/5}}{(1+c\phi)^{6/5}+\mu(2\beta^{2})^{-1}\phi^{6/5}}
\end{equation}
It is an attractive feature of our theory that similar potentials can
describe the nonsingular cosmological evolution of a universe with
and without matter.
We cannot solve for the time evolution analytically with this potential,
so we evolve (14) and (39) numerically with a Runge-Kutta algorithm
for $a(t)$ and $D(t)$, using the fact that in two dimensions with our
ansatz for the metric, the Ricci scalar takes the simple form
$R=2{\ddot a}/a$. In Figures 5,6 and 7 we plot the scale factor $a(t)$ ,
the Ricci scalar $R(t)$ and the dilaton $\phi(t)$ for
$\mu=2\beta^{2}=1$ and $\kappa^{2}=0.1$ .
The evolution from the de Sitter bounce into a matter dominated
FRW universe can clearly be seen.

Our spacetime has finite curvature everywhere for $t \geq 0$ . It can be
made geodesically complete by patching on the contracting part of the de Sitter
bounce for negative times. This is possible since the matching conditions for
the metric are
satisfied at $t=0$, namely $a(t)$ and $\dot a(t)$ are continuous. Also,
$\phi(t)$ is finite at $t=0$ and does not cause any problems. Therefore we have
indeed obtained a nonsingular two-dimensional cosmological model.

\section{Conclusion}
Since quantum gravity is non-renormalizable\cite{H&V 74} and at present we can
extract little information from string theory concerning the quantum nature of
spacetime, we have used a toy model in 1+1 dimensions to investigate what
conclusions may be reached at the classical level concerning singularities. The
key assumption in our work is that the quantum processes responsible for the
removal of singularities may be represented in an effective, classical theory
of gravity, valid at curvature scales intermediate between those of any future
theory of everything and GR.

Motivated by the low energy limit of string theory and attempts to apply these
ideas in four spacetime dimensions\cite{M&B 92} we have constructed a class of
two dimensional models for gravity for which all homogeneous and isotropic
solutions are nonsingular and approach a FRW matter dominated universe at late
times. We have also demonstrated that our construction is valid for both the
vacuum theory and, for a large class of initial conditions, when a conformally
coupled scalar field is included in the model.

Our construction is based on explicitly limiting the Ricci scalar (the only
curvature invariant in two dimensions) by means of a non-dynamical scalar field
$\phi$ which we refer to as the dilaton. The particular choice of potential for
the dilaton defines the theory and ensures that the solutions remain
singularity free as regions of high curvature are approached.

If the type of behaviour that we reveal here is realized by nature, there may
be very interesting implications for cosmology. The existence of a de Sitter
phase arising purely from the gravitational sector at early times (also seen in
the four dimensional case\cite{M&B 92}) may provide us with a natural
inflationary epoch. Although at present inflation appears to be the only viable
solution to the horizon problem, there are difficulties associated with the
fine tuning of parameters neccessary to implement this scenario using
elementary scalar fields. Thus, an alternative formulation such as ours may be
very attractive for cosmology.

\vspace{1cm}
\begin{center}
\bf Acknowledgements
\end{center}
\vspace{5mm}
We would like to thank Robert Brandenberger, Slava Mukhanov and Andrew
Sornborger for suggestions and helpful discussions. This work was supported in
part by the US Department of Energy under Grant DE-FG0291ER40688.

\newpage
\begin{center}
\bf Figure Captions
\end{center}
Figure 1: The scale factor of the universe, as a function of the dilaton
coupling, for the vacuum theory with an expanding de Sitter solution.
\vspace{5mm}
\newline
Figure 2: An example of the scale factor, as a function of cosmological time,
in the vacuum theory with an expanding de Sitter solution.
\vspace{5mm}
\newline
Figure 3: The Ricci scalar, as a function of time, for the vacuum solution with
the same parameters as those used in Figure 2.
\vspace{5mm}
\newline
Figure 4: The dilaton, as a function of time, for the vacuum solution of Figure
2.
\vspace{5mm}
\newline
Figure 5:  An example of the scale factor, as a function of cosmological time,
when a conformally coupled scalar field has been included. This example shows a
de Sitter bounce.
\vspace{5mm}
\newline
Figure 6: The Ricci scalar, as a function of time, for the matter solution of
Figure 5.
\vspace{5mm}
\newline
Figure 7: The dilaton, as a function of time, for the matter solution of Figure
5.
\end{document}